\documentclass[twocolumn,aps]{revtex4-1}

\usepackage{graphicx}
\usepackage{amssymb}
\usepackage{amsfonts}
\usepackage{amsmath}
\usepackage{amsbsy}
\usepackage{natbib}

\usepackage{color}
\usepackage{float}

\renewcommand{\vec}[1]{\mbox{\boldmath$\mathrm{#1}$}}
\newcommand{\be}{\begin{equation}}
\newcommand{\ee}{\end{equation}}
\newcommand{\ben}{\begin{eqnarray}}
\newcommand{\een}{\end{eqnarray}}

\begin{document}

\title{Stratonovich-Ito integration scheme in ultrafast spin caloritronics}

\author{L. Chotorlishvili$^1$, Z. Toklikishvili$^2$, X.-G. Wang$^{3}$, V.K. Dugaev$^{4}$, J. Barna\'{s}$^{5,6}$ J. Berakdar$^1$}

\address{$^1$ Institut f\"ur Physik, Martin-Luther Universit\"at Halle-Wittenberg, D-06120 Halle/Saale, Germany \\
$^2$  Faculty of Exact and Natural Sciences, Tbilisi State University, Chavchavadze av.3, 0128 Tbilisi, Georgia\\
$^3$ School of Physics and Electronics, Central South University, Changsha 410083, China\\
$^{4}$ Department of Physics and Medical Engineering, Rzeszow University of Technology, 35-959 Rzeszow, Poland\\
$^{5}$Faculty of Physics, Adam Mickiewicz University, ul. Umultowska 85, 61-614 Poznan, Poland\\
$^{6}$Institute of Molecular Physics, Polish Academy of Sciences, ul. M. Smoluchowskiego 17, 60-179 Pozna\'{n}, Poland}

\begin{abstract}
The  magnonic spin Seebeck effect  is a key element of spin caloritronic, a field that exploits thermal effects for spintronic applications.
 Early studies were focused on investigating  the steady-state nonequilibrium magnonic spin Seebeck current, and
  the underlying physics  of the magnonic spin Seebeck effect is now relatively well established. However,  the initial steps of the formation of the spin Seebeck current are in the scope of recent interest. To address this dynamical aspect theoretically  we propose here a new approach to the time-resolved spin Seebeck effect. Our method exploits the supersymmetric theory of stochastics and Ito - Stratonovich integration scheme. We found that in the early step the spin Seebeck current has both nonzero transversal and longitudinal components. As the magnetization dynamics approaches the steady-state, the transversal components decay through  dephasing over the dipole-dipole reservoir. The time scale for this process is typically  in the sub-nanoseconds pointing thus to the potential of an ultrafast control of the dynamical spin Seebeck during its buildup.
\end{abstract}
\date{\today}
\maketitle

\section{Introduction}

Irradiating magnetic samples with electromagnetic fields  may result in a variety of phenomena, including
subpicosecond magnetic order breakdown,   electron-phonon spin-flip scattering \cite{Koopmans}, electron-magnon scattering in  non-equilibrium   \cite{Tserkovnyak}, and superdiffusive spin transport  \cite{Battiato}. These observations are related to ultrafast spin dynamics \cite{Chubykalo-Fesenko, Atxitia} and depend on the parameters of the driving fields such as their intensity, duration, and  frequencies, we well as on  the inherent properties of the magnetic sample.
 Our interest here is devoted to a particular aspect namely to the non-equilibrium magnonic current generated by a temperature gradient due to local  heating of the sample by a laser pulse \cite{xiguangfront}. This means that we concentrate on the regime where the  phonon temperature profile has already been established and consider the nonequilibrium dynamics of magnons. We note that magnon dynamics is of a particular importance for applications, as magnons  are low energy excitations that can carry information over long distances and can be utilized for logic operations. To deal with  non-equilibrium processes under the influence of irregular forces and thermal fluctuations the  Fokker–Planck (FP) equation is the method of choice \cite{H. Risken,W. T. Coffey,S. V. Titov,D. A. Garanin1,D. A. Garanin2,miyazaki98}.

In general, FP applies also to nonlinear (chaotic)  systems with positive Lyapunov exponents \cite{Schwab}.
Treating the thermally activated magnetization dynamics and the steady-state magnonic spin current,  the FP equation allows obtaining results beyond the linear response theory \cite{Chotorlishvili2013,Chotorlishvili2019}. However, the corresponding  nonstationary case has not yet been treated with FP equation. In fact, FP equation is a nonlinear partial differential equation which admits  exact analytical time-dependent solution only in a few limited cases.
Our aim to describe the ultrafast spin dynamics entails  access to the time-dependent solution of the FP equation. The available procedures and analytical tools for solving the time-dependent FP equation are  limited   basically  to 1D systems.  As an alternative, one can consider a supersymmetric theory of stochastics and the Stratonovich-Ito integration scheme.
In this work, we apply the Stratonovich-Ito integration scheme to the system below the Curie temperature.

Here we  present  an analytical FP-based approach to  study of thermally activated ultrafast magnonic spin current. Specifically,  we focus on the behavior of the non-equilibrium  spin current, generated at the interface of ferromagnetic insulator and normal metal \cite{XiaoBauerUchida,K. I. Uchida} and calculate how it approaches  the nonequilibrium (steady) state. To this end the evaluation of the correlation functions in the nonequilibrium state is needed, and as we show here, this can be achieved  by using the FP equation and the Stratonovich-Ito integration scheme for the stochastic noise.

Our choice of the sample is motivated by the recent experiments uncovering  the early stage  of the spin Seebeck effect \cite{P. W. Brouwer}. Using terahertz spectroscopy applied to   bilayers of ferrimagnetic yttrium iron garnet (YIG) and platinum, the spin Seebeck current is shown to arise on the  $\sim100$ fs time scale.

\begin{figure}[htbp]\label{pictorial}
	\includegraphics[width=0.48\textwidth]{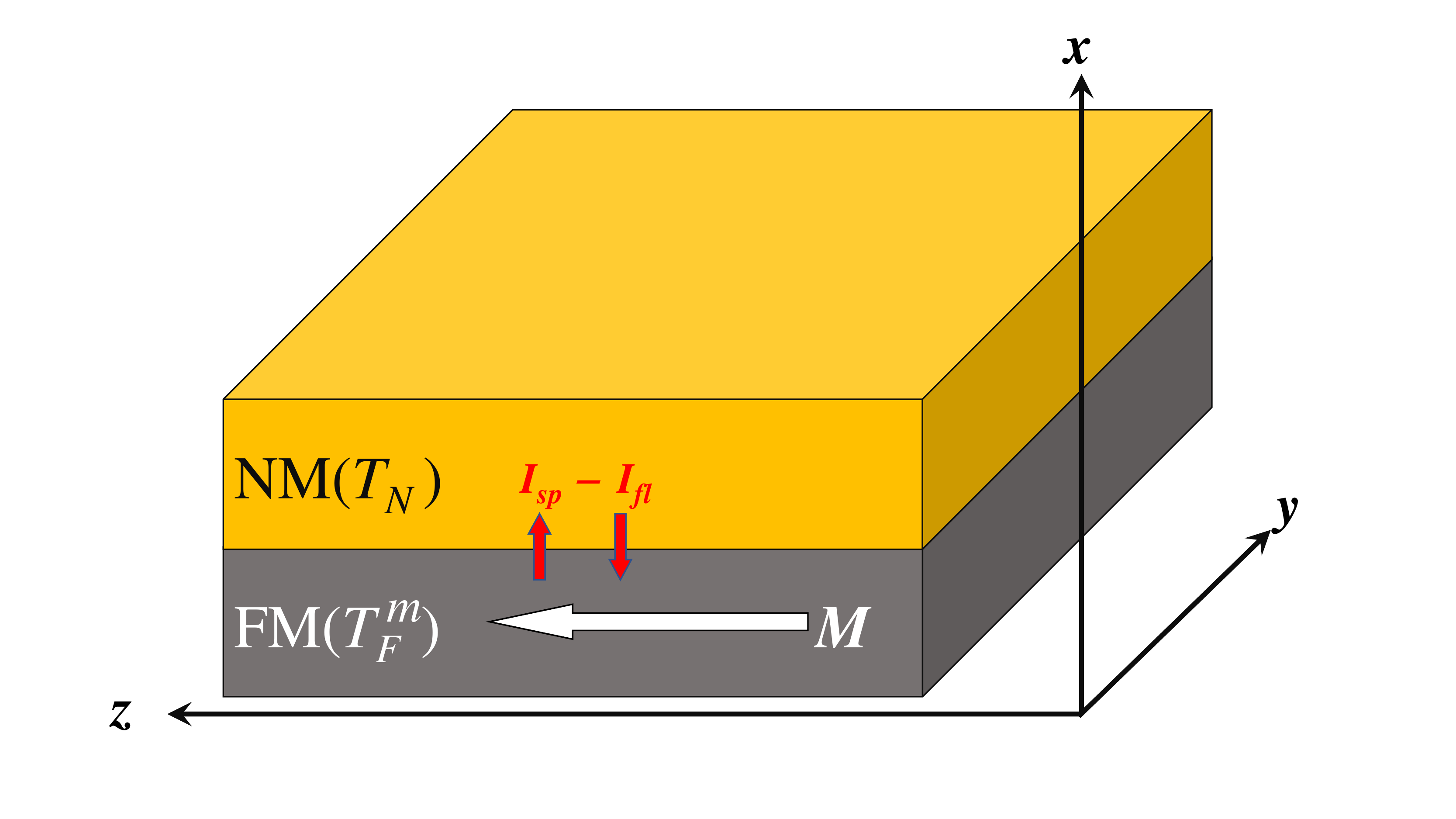}
	\caption{The schematics of the system. The red arrows show the direction of spin pumping $I_{sp}$ and fluctuating  $I_{fl}$ spin currents flowing from ferromagnetic insulator to normal metal ($I_{sp}$) and from normal metal to ferromagnetic insulator ($I_{fl}$). Equilibrium magnetization in the magnetic insulator is along the $\vec{Z}$ axis. $ T_{F}^{m} $ is the magnon temperature in the magnetic insulator, and $T_N$ is the temperature of the normal metal.}
\end{figure}

The work is organized as follows: In section \textbf{II}, we define the magnonic spin current. In section \textbf{III}, we describe the theoretical methods used afterward. In section \textbf{IV} we present results and conclude the work. 

\section{Modelling non-equilibrium magnonic spin current}

The total spin current $\vec{I_{tot}}=\vec{I}_{sp}+\vec{I}_{fl}$ crossing the normal-metal/ferromagnet interface has two contribution: spin pumping current $\vec{I}_{sp}$ flowing from the ferromagnetic insulator to the normal metal and the fluctuating spin current $\vec{I}_{fl}$ flowing in the reverse direction.
Equilibration of electronic and phononic degrees proceeds much faster (subpicoseconds) than magnons (up to nanoseconds). As we are interested in the dynamics of the latter we assume that the  temperature $T_{F}$ in the ferromagnetic  layer   is set by the equilibrium  electrons and phonon temperature. The same applies to the  temperature $T_{N}$ in the normal metal. The heating is assumed to be induced by a laser pulse. The relation of the pulse parameters and the value  of the temperature has been discussed in details in Ref.\cite{xiguangfront}.
 The thermal bias through the mismatch between magnon temperature $T_{F}^{m}$ and the sample temperatures $T_{F}$ drives  the  magnonic spin current of interest here. We note,
 magnons are low-energy elementary excitations of the ordered phase. Thus, spin (or electron/lattice) dynamics at (fs) times where the magnetic state is broken down or not yet established  is not discussed here. \\
The spin pumping current flowing from the ferromagnetic insulator into the normal metal reads
\cite{Tserkovnyak1,Tserkovnyak2,AdachiUchidaSaitohMaekawa}
\begin{equation}\label{spinpump}
\vec{I}_{sp}(t)=\frac{\hbar}{4\pi}[g_{r}\vec{m}(t)\times \dot{\vec{m}}(t)+g_{i}\dot{\vec{m}}(t)],
\end{equation}
where $g_{r}$ and $g_{i}$ are the real and imaginary parts of the dimensionless spin mixing conductance of the ferromagnet/normal-metal $(F|N)$ interface, while $\vec{m}(t)=\vec{M}(t)/M_s$ is the dimensionless unit vector along the magnetization orientation (here $M_{s}$ is the saturation magnetization) and $\dot{\vec{m}}\equiv d\vec{m}/dt$.
The spin current is a tensor object characterized by the direction of the current flow and the orientation of the flowing spin (magnetic moment).
Due to the geometry of the system, see \cite{XiaoBauerUchida}, the pumping spin current flows along the $\vec{z}$-axis while the fluctuating spin current flows in the opposite ($-\vec{z}$) direction,
\begin{equation}\label{NouseSpinCurrent}
\vec{I}_{fl}(t)=-\frac{M_{s}V}{\gamma}\vec{m}(t)\times\vec{\zeta}'(t).
\end{equation}
Here, $V$ is the total volume of the ferromagnet, $\gamma$ is the gyromagnetic factor, and
$\vec{\zeta}'(t)=\gamma\vec{h}'(t)$ with $\vec{h}'(t)$ denoting the random magnetic field. In the classical limit, $k_{B}T\gg\hbar\omega_{0}$, the correlation function $\langle \zeta'_{i}(t)\zeta'_{j}(t')\rangle$  of
$\vec{\zeta}'(t)$ reads
\begin{equation}\label{correlation1}
\langle \zeta'_{i}(t)\zeta'_{j}(0)\rangle =\frac{2\alpha'\gamma k_{B}T_{N}}{M_{s}V}\delta_{ij}\delta(t)\equiv \sigma'^2\delta_{ij}\delta(t)
\end{equation}
for $i,j=x,y,z$, where $\langle...\rangle$ denotes the ensemble average,
$\omega_{0}$ is the ferromagnetic resonance frequency, and $\alpha'$ is the contribution to the damping constant due to  spin pumping, $\alpha^\prime =\gamma\hbar g_{r}/4\pi M_{s}V$.  We note that the correlator  (Eq.(\ref{correlation1})) is proportional to the temperature of the normal metal $T_{N}$.
The total spin current thus reads
\begin{equation}\label{totalspincurrent}
\langle\vec{I}_{tot}\rangle=\frac{M_{s}V}{\gamma}[\alpha'\langle\vec{m}\times\dot{\vec{m}}\rangle-\langle\vec{m}\times\vec{\zeta'}\rangle].
\end{equation}

The  temperature-dependent magnetization dynamics is governed by the stochastic LLG equation
\begin{equation}\label{LLG}
\frac{d\vec{m}}{dt}=-\gamma\vec{m}\times(\vec{H}_{\rm eff}+\vec{h})+\alpha\vec{m}\times\dot{\vec{m}},
\end{equation}
where $\alpha$ is the Gilbert damping constant, $\vec{H}_{\rm eff}$ is the effective field, and the time-dependent random magnetic field in the ferromagnet is described by $\vec{h}$.
This effective field $\vec{H}_{\rm eff}$ contains the anisotropy field $H_{A}$ and the external magnetic field $H_{0z}$ oriented along the $\vec{z}$-axis.
The total random magnetic field $\vec{h}(t)$ has two contributions from independent noise sources: the thermal random field $\vec{h}_{0}(t)$,
and the random field $\vec{h}'(t)$. The former is related to the finite temperature in the ferromagnetic insulator and the second to the fluctuations in the normal metal.
The correlators of the statistically independent noise sources are additive leading to the effective (enhanced) magnetic damping constant $\alpha=\alpha_{0}+\alpha'$ \cite{XiaoBauerUchida} ($\alpha_0$ is the damping parameter of the ferromagnetic material, meaning  without the contributions from the pumping currents),
\begin{equation}\label{correlation2}
\langle\zeta_{i}(t)\zeta_{j}(0)\rangle=\frac{2\alpha\gamma k_{B}T_{F}^{m}}{M_{s}V}\delta_{ij}\delta(t)=\sigma^{2}\delta_{ij}\delta(t),
\end{equation}
where $\vec{\zeta}(t)=\gamma\vec{h}(t)$, and $\alpha T_{F}^{m}=\alpha_{0}T_{F}+\alpha' T_{N}$.
\section{Theoretical method}
To find the total spin current (Eq.(4)) we  use the FP equation for the distribution function of the magnetization $P(m_z,t)$ which is related to the stochastic equation of the magnetic dynamics (Eq.(5))
\begin{equation}
\frac{\partial P(m_z,t)}{\partial t}
=\frac{\partial }{\partial m_z}\Big[ \frac{\partial }{\partial m_z}
+\beta U'(m_z)\Big] P(m_z,t),
\end{equation}
where $U(m_{z})=2\alpha\left(\omega_{0}m_{z}-\frac{\omega_{p}m_{z}^{2}}{2}\right)$ is the potential, $\beta =1/\sigma^{2}$ is the effective inverse temperature, $\omega_{p}=\gamma H_{A}$, and $H_{A}$ is the anisotropy field. As detailed above, in our case $U(m_{z})$ is time-independent.
The stationary solution of Eq.~(7) is given by \cite{Chotorlishvili2013}
\begin{eqnarray}\label{SUSY0}
&&P_0(m_{z})=Z^{-1}\exp\left[-\beta U(m_{z})\right],\nonumber\\
&&Z=\int\exp\left[-\beta U(m_{z})\right]d^{3}\vec{m}.
\end{eqnarray}
For the time-dependent distribution one makes the Ansatz
\begin{equation}\label{SUSY1}
P(m_{z},t)=\psi(m_{z},t)\, \exp \Big[ -\frac{\beta }2\, U(m_{z})\Big] .
\end{equation}
Using Eq.(9) and Eq.(7) we find that $\psi(m_{z},t)$ is a solution of the Schr\"odinger equation for imaginary time,
\begin{eqnarray}\label{SUSY2}
\frac{\partial \psi(m_{z},t)}{\partial t}=-\hat{H}\, \psi(m_{z},t),
\end{eqnarray}
with the Hamiltonian
\begin{eqnarray}
\label{11}
\hat{H}=-\frac{d^{2}}{dm_z^{2}}+\left(\frac{U'(m_z)}{2\sigma^{2}}\right)^{2}
-\frac{U''(m_z)}{2\sigma^{2}}.
\end{eqnarray}
As  $U$ is time independent, the general solution of Eq.~(11) is  $\psi(m_{z},t)=\sum\limits_{n}C_{n}\exp\left(-\lambda_{n}t\right)\psi_{n}(m_{z})$,
where $\psi _n(m_z)$ and $\lambda _n$ are the eigenfunctions and eigenvalues of the stationary equation, $\hat{H}\psi_{n}(m_{z})=\lambda_{n}\psi_{n}(m_{z})$.
Hence, the problem of solution of the time-dependent FP equation reduces to the determination of $\lambda _n$ and the corresponding eigenfunctions of the Hamiltonian $\hat{H}$. The calculations can be substantially simplified due to the hidden supersymmetry of this problem \cite{tsvelik}. Indeed, one can introduce the supersymmetric Hamiltonian (for the supersymmetry see \cite{Melnikov, Witten, BernsteinBrown,Parisi,Sourlas,CaroliCaroli}) $\hat{H}_{\rm susy}=Q^{\dag}Q+QQ^{\dag}={\rm diag}\, (\hat{H}_{+},\hat{H}_{-})$, where
\begin{equation}\label{SUSY3}
Q=\left[ {\begin{array}{cc}
	0 & 0 \\ A & 0 \\
	\end{array} } \right], \;
Q^{\dag}=\left[ {\begin{array}{cc}
	0 & A^{\dag} \\	0 & 0 \\
	\end{array} } \right]
\end{equation}
and $A=\frac{1}{2}\beta U'(m_z)-\partial/\partial m_z$. Note that $Q$ and $Q^\dag $ are  nilpotent operators, $Q^2=(Q^\dag )^2=0$, and the commutator $[Q, \hat{H}_{susy}]=0$. As a result, $\hat{H}_+$ and $\hat{H}_-$ have common eigenfunctions: if $\Psi _n$ is an eigenfunction of $\hat{H}_+$ then $Q\Psi _n$ is the eigenfunction of $\hat{H}_-$ (except for the ground state corresponding to $\lambda _0=0$).
The operator $\hat{H}_{\rm susy}$  acts in the space of Bose and Fermi fields. Namely, $\hat{H}_+=d^2/dm_z^2+V_+(m_z)$ is the operator for bosons and $\hat{H}_-=d^2/dm_z^2+V_-(m_z)$ for fermions, where $V_\pm (m_z)=(U'/2\sigma ^2)^2\pm U''/2\sigma ^2$ are the corresponding potentials. The operator $Q$ transforms bosons to fermions and vice versa.

 $\hat{H}_+$ coincides with Hamiltonian (11). It is however more convenient  to solve the Schr\"odinger equation with fermionic Hamiltonian $\hat{H}_-$ because the corresponding potential $V_-(m_z)$ is close to the parabolic form. Owing to the supersymmetry, the eigenfunctions and eigenvalues are the same. Using this approach we find $\lambda_{1}=\frac{\sigma^{2}}{2\pi}\exp\left(-\alpha\omega_{p}/\sigma^{2}\right)$,
and in the limit of strong anisotropy
\begin{equation}\label{SUSY4}
\lambda_{n}\approx 4\alpha \omega_{p}\left(n-1\right)/\sigma^{2}.
\end{equation}
The first non-vanishing $1/\lambda_{2}$ defines the characteristic relaxation time scale. For more details on the supersymmetry theory of stochastics  we refer to \cite{Girardello, Cecotti, Horgan, Kleinert, Khomenko, Dijkgraaf, Ovchinnikov}.

To explore the time dependence of the non-equilibrium spin current $\langle\vec{I}(t)_{tot}\rangle$ we utilize the Stratonovich-Ito integration scheme
\cite{Kunita, Baxendale, Gardiner} and construct a reductive perturbation theory valid in the low-temperature limit (specified below).
We briefly recall the main concepts of the stochastic Ito - Stratonovich integration. The time integral from the stochastic noise is equal to the function $W(t)$ which has no time derivative ($W(t)$ is not a smooth function) $\int\limits_{0}^{t}\xi(\tau)d\tau =W(t)$. Therefore, the stochastic integration is  performed  using  the mean-square (ms) convergence of the
sequence of the random variable $X_{n}(\omega)$, meaning that
\begin{eqnarray}\label{Ito - Stratonovich1}
ms\left\{\lim\limits_{n\rightarrow\infty} X_{n}\right\}=X,
\end{eqnarray}
is equivalent to
\begin{widetext}
\begin{eqnarray}\label{Ito - Stratonovich2}
\lim\limits_{n\rightarrow\infty}\int\limits_{-\infty}^{\infty}p(\omega)\left[X_{n}(\omega)-X(\omega)\right]^{2}=
ms \left\{\lim\limits_{n\rightarrow\infty}\left\langle\left(X_{n}-X\right)^{2}\right\rangle\right\}=0.
\end{eqnarray}
\end{widetext}
Here $p(\omega)$ is the probability distribution function.
The stochastic integral is defined as follows:
\begin{widetext}
\begin{eqnarray}\label{Ito - Stratonovich3}
\int\limits_{t_{0}}^{t}G(\tau)\, dW(\tau)=ms\left\{\lim\limits_{n\rightarrow\infty} \sum\limits_{i=1}^{n}G(t_{i-1})\left[W(t_{i})-W(t_{i-1})\right]\right\},
\end{eqnarray}
\end{widetext}
where $G(t)$ is an arbitrary function of time.
We assume that the magnon temperature in the system is low, which means that  the thermal energy is smaller than the anisotropy barrier. Therefore, the appropriate ansatz for the solution of the stochastic LLG equation is
\begin{eqnarray}\label{Ito - Stratonovich4}
\vec{m}(t)=\vec{m}_{0}(t)+\varepsilon \vec{m}_{1}(t),
\end{eqnarray}
where $\vec{m}_{0}(t)$ is the deterministic solution and $\vec{m}_{1}(t)$ is the correction due to the stochastic field.
The equation for the stochastic part reads
\begin{eqnarray}\label{Ito - Stratonovich5}
&& d\vec{m}_{1}(t)=-A\left[ \vec{m}_0(t)\right] \, \vec{m}_{1}(t)\, dt+\nonumber\\
&& B\left[ \vec{m}_0(t)\right] \, d\vec{W}(t),
\end{eqnarray}
where $d\vec{W}(t)=\vec{\xi}(t)dt$  and for brevity we introduced the notations
\begin{widetext}
\begin{equation}\label{Ito - Stratonovich6}
   A\left[ \vec{m}_0(t)\right]=
  \left[ {\begin{array}{ccc}
   0 & \omega_{\rm eff}(t) & 0\\
   -\omega_{\rm eff}(t) & 0 & 0\\
   A & 0 & 0\\
  \end{array} } \right],~~~
   B\left[ \vec{m}_0(t)\right]=
  \left[ {\begin{array}{ccc}
   0 & m_{0z}(t) & -m_{0y}(t)\\
   -m_{0z}(t) & 0 & m_{0x}(t)\\
   m_{0y}(t) & -m_{0x}(t) & 0\\
  \end{array} } \right],
\end{equation}
\end{widetext}
with $\omega_{\rm eff}\left(\omega_{0}+\omega_{p}m_{z}\right)$.
Taking into account Eq.(\ref{Ito - Stratonovich1})-Eq.(\ref{Ito - Stratonovich6}),
after relatively  involved analytical calculations for the correlation functions and the non-equilibrium spin current we deduce
\begin{eqnarray}\label{Ito - Stratonovich7}
&&\left\langle\vec{I}_{s}(t)\right\rangle=2\alpha'k_{B}\varepsilon^{2}\vec{m}_{0}(t)\left(T_{F}^{m}-T_{N}\right),\nonumber\\
&&\left\langle m_{1i}(t)\xi_{j}(t)\right\rangle=\sigma^{2}\varepsilon_{ijk}m_{0k}(t),\nonumber\\
&&\left\langle m_{1i}(t)\xi'_{j}(t)\right\rangle=\sigma'^{2}\varepsilon_{ijk}m_{0k}(t).
\end{eqnarray}
In the case of a weak anisotropy Eq.(\ref{Ito - Stratonovich7}) simplifies and for the non-equilibrium magnonic spin current components we obtain:
\begin{eqnarray}\label{Ito - Stratonovich8}
&&\left\langle I^{x}_{s}(t)\right\rangle=2\alpha'k_{B}\varepsilon^{2}\frac{\cos(\varphi_{0}+\omega_{0}t)}{\cosh \alpha \omega_{0}t}\left(T_{F}^{m}-T_{N}\right),\nonumber\\
&&\left\langle I^{y}_{s}(t)\right\rangle=2\alpha'k_{B}\varepsilon^{2}\frac{\sin(\varphi_{0}+\omega_{0}t)}{\cosh \alpha \omega_{0}t}\left(T_{F}^{m}-T_{N}\right),\nonumber\\
&&\left\langle I^{z}_{s}(t)\right\rangle=2\alpha'k_{B}\varepsilon^{2}\tanh \left(\alpha \omega_{0}t\right)\left(T_{F}^{m}-T_{N}\right).
\end{eqnarray}
From Eq.(\ref{Ito - Stratonovich8}) follows that in the asymptotic, long-time limit the
only component of the magnonic spin current that survives is $\left\langle I^{z}_{s}(t)\right\rangle$, and we recover  the classical result of Xiao \textit{et al} \cite{XiaoBauerUchida}. For short times, however, the other components are sizable and even dominant and thus can thus be exploited for ultrafast picosecond magnonics.

We applied the Stratonovich-Ito integration scheme to the system below the Curie temperature. Nevertheless, our method can be extended to the Landau-Lifshitz-Bloch equation as well. Note that Eq.(\ref{Ito - Stratonovich5}), in the coefficient $A\left[ \vec{m}_0(t)\right] $ and $B\left[ \vec{m}_0(t)\right] $, contains the solution of the deterministic Landau-Lifshitz-Gilbert equation. One can replace the solution of the deterministic LLG equation $\vec{m}_0(t)$ by the solution of the deterministic Landau-Lifshitz-Bloch equation with extra longitudinal damping parameter \cite{OstlerHinzkeNowak}. After this replacement, we can again perform Stratonovich-Ito integration.

The result Eq.(\ref{Ito - Stratonovich8}) is obtained in the single macrospin approximation but can be  generalized to an extended system using the ensemble averaging over the dipole-diploe reservoir. We note that the transversal spin current components in Eq.(\ref{Ito - Stratonovich8}) contain the rotating terms. In the case of  extended systems, each spin rotates with a slightly different frequency due to the broadening of the resonance frequency $\omega_{0}$. Precession with different frequencies leads to the dephasing of the signal in time. We assume that the dephasing of the transversal magnetization and current components have the same nature. Following \cite{AbragamGoldman}, we write down the equation for the transversal magnetization component
\begin{eqnarray}\label{AbragamGoldman1}
-i\hbar\frac{dm_{x}(t)}{dt}=\left[\hat{H}_{d}(t),m_{x}(t) \right],
\end{eqnarray}
or in the matrix form
\begin{eqnarray}\label{AbragamGoldman2}
-i\hbar\frac{d(m_{x}(t))_{nn'}}{dt}=\hbar \Delta \omega(t)_{nn'}(m_{x}(t))_{nn'}.
\end{eqnarray}
The Hamiltonian $\hat{H}_{d}(t)$ in Eq.(\ref{AbragamGoldman1}), (\ref{AbragamGoldman2}) describes the dipole-dipole reservoir, and the time dependence of the Heisenberg operators is governed through the
Zeeman Hamiltonian $\hat{H}_{Z}$, (see \cite{AbragamGoldman} for more details).
Let us quantify the fluctuations of the local field through the function
\begin{eqnarray}\label{AbragamGoldman3}
\left\langle \Delta \omega(t)_{nn'} \Delta \omega(t+\tau)_{nn'}\right\rangle=M_{2}\Psi(\tau),
\end{eqnarray}
where
\begin{eqnarray}\label{AbragamGoldman4}
M_{2}=-\frac{Tr\left\{\left[\hat{H}_{d},m_{x}\right]^{2}\right\}}{\hbar^{2}Tr\left\{m_{x}^{2}\right\}}-\omega_{0}^{2},
\end{eqnarray}
is the second moment of the transversal component. We assume that the dephasing mechanism of the transversal spin current components is the same. Taking into account Eq.(\ref{AbragamGoldman1})-Eq.(\ref{AbragamGoldman4}) for the ensemble averaged dephasing transversal spin currents we infer
\begin{widetext}
\begin{eqnarray}\label{AbragamGoldman5}
&&\left\langle \langle I^{x}_{s}(t)\right\rangle\rangle=2\alpha'k_{B}\varepsilon^{2}\frac{\cos(\varphi_{0}+\omega_{0}t)}{\cosh \alpha \omega_{0}t}\exp\left[-M_{2}\int\limits_{0}^{t}(t-\tau)\Psi(\tau)d\tau\right]\left(T_{F}^{m}-T_{N}\right),\nonumber\\
&&\left\langle\langle I^{y}_{s}(t)\right\rangle\rangle=2\alpha'k_{B}\varepsilon^{2}\frac{\sin(\varphi_{0}+\omega_{0}t)}{\cosh \alpha \omega_{0}t}\exp\left[-M_{2}\int\limits_{0}^{t}(t-\tau)\Psi(\tau)d\tau\right]\left(T_{F}^{m}-T_{N}\right).
\end{eqnarray}
\end{widetext}
In the limit of the white noise the dephasing exponent takes the simpler form:
$\left\langle \langle I^{x,y}_{s}(t)\right\rangle\rangle\approx \left\langle \langle I^{x,y}_{s}(0)\right\rangle\rangle\exp\left[-t/T_{2}\right]$, where the transversal relaxation time is given by $T_{2}=-M_{2}t\int\limits_{0}^{\infty}\Psi(\tau)d\tau$.
Thus, the decay of the transversal magnonic spin current components in the ultrafast spin Seebeck effect is solely determined by the dipole-dipole interactions.

\section{Results and discussions}

\begin{figure}[htbp]
	\includegraphics[width=0.48\textwidth]{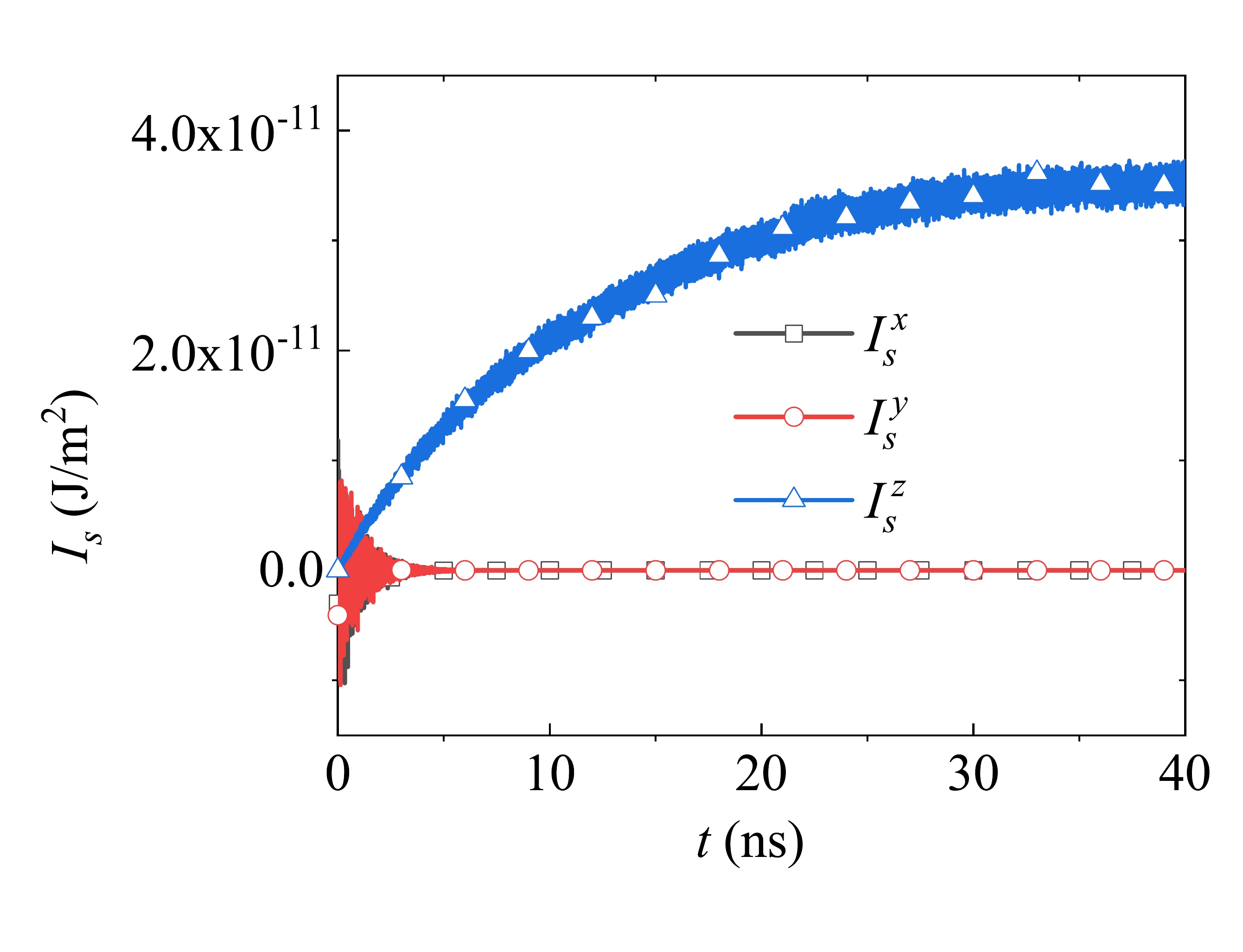}
	\caption{\label{figure1} Time dependent non-equilibrium transversal and longitudinal magnonic spin current components $ I^x_s(t) $, $ I^y_s(t) $ and $ I^z_s(t) $. The magnon temperature is equal to $ T_{F}^{m} = 5 $ K and temperature of the normal metal $ T_N = 0 $.  The external magnetic field $H_{0z} = 2 \times 10^5$ A/m is applied in $ +z $ direction.}
\end{figure}

\begin{figure}[htbp]
	\includegraphics[width=0.48\textwidth]{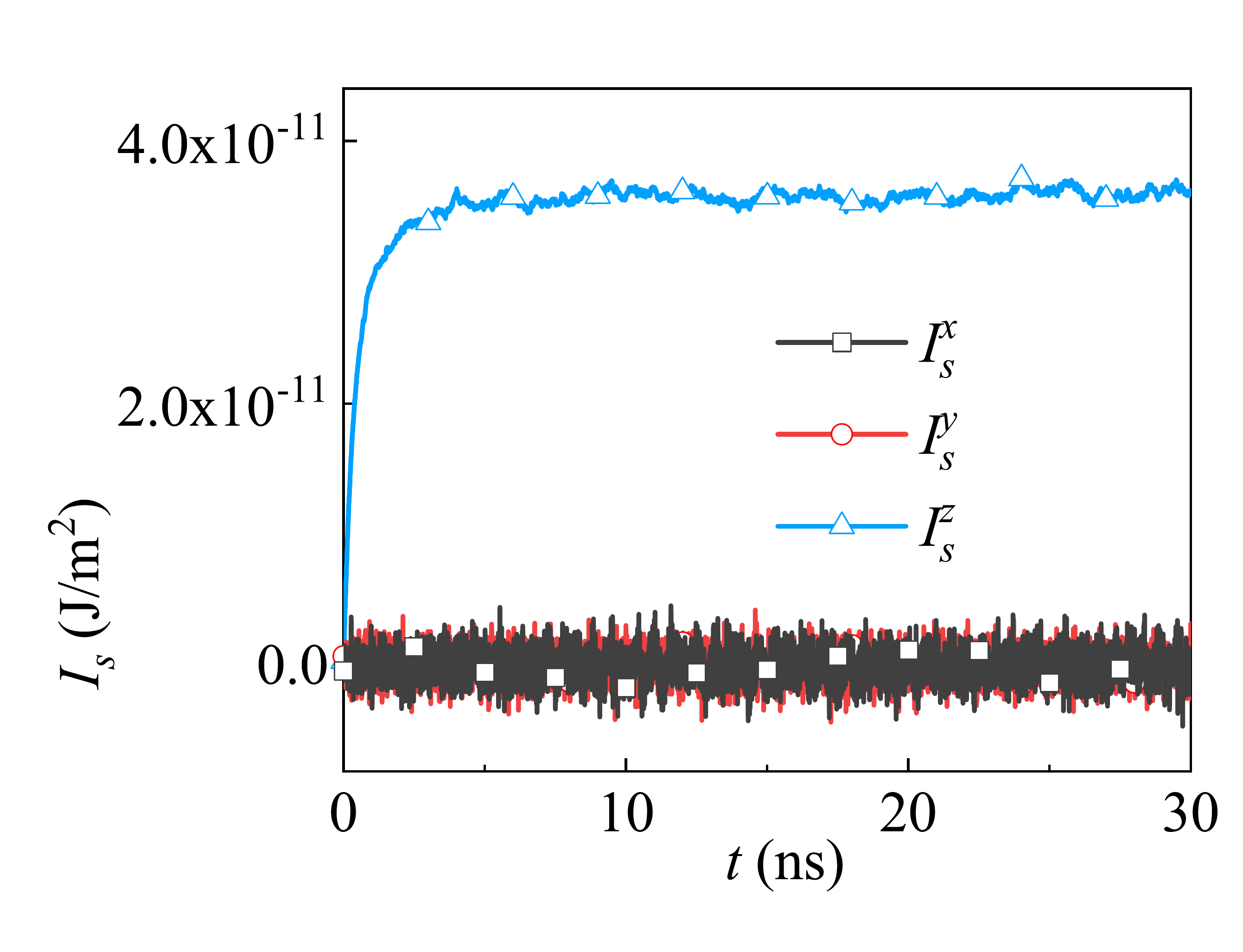}
	\caption{\label{figure2}  Time dependent, non-equilibrium transversal and longitudinal magnonic spin current components $ I^x_s(t) $, $ I^y_s(t) $ and $ I^z_s(t) $ for a finite ferromagnetic sample. The short time scale longitudinal magnonic spin currents are  important,  whereas in the long-time limit  $ I^z_s(t) $ is dominant.}
\end{figure}

In the numerical simulation, the motion of $ \vec{m} $ is governed by the LLG equation (\ref{LLG}). The adopted numerical parameters are $ M_s = 1.4 \times 10^5 $ A/m, the damping constant $ \alpha = 0.001 $, the external magnetic field $H_{0z} = 2 \times 10^5$ A/m, and the spin-mixing conductance $ g_r = 3 \times 10^{15} $ 1/m$ ^2 $. In the equilibrium state, the local magnetization points along the $ +z $ direction. We set the temperature $ T_{F}^{m} = 5K $ and $ T_N = 0 $. The time-dependent magnonic spin pumping currents $ I^x_s(t) $, $ I^y_s(t) $ and $ I^z_s(t) $ are plotted in Fig.\ref{figure1}. The spin current is calculated  using  Eq. (\ref{spinpump}). For the transversal components $ I^x_s(t) $ and  $ I^y_s(t) $ we consider  the averaging procedure through the exponential factors $\langle\langle I^x_s(t)\rangle\rangle = e^{-t/T_{2}} \langle I^x_{s,0} \rangle$ and $\langle\langle I^y_s(t) \rangle\rangle= e^{-t/T_{2}}\langle I^y_{s,0}\rangle$, with the transversal relaxation time $T=N/\omega_{0}$, $ N = 50 $, and the values of $ \langle I^x_{s,0}\rangle $ and $ \langle I^y_{s,0}\rangle $ are calculated from Eq. (\ref{spinpump}). The numerical solution plotted in Fig. \ref{figure1} is in good agreement with the analytical results expressed by Eqs. (\ref{Ito - Stratonovich7}) and (\ref{AbragamGoldman5}). Calculations done for the anisotropy field $ H_z = \frac{2 K_z}{\mu_0 M_s} $ (along $ z $ axis) with constant $ K_z = 1.8 \times 10^4 $ J/m$ ^3 $ (not shown) leads to  similar conclusions.
To explore the dephasing problem for an extended sample, we performed  micromagnetic simulations. The results of the simulations are presented in the supplementary information.
We performed  numerical simulations for an extended ferromagnetic sample. Our simulations include the effects of the dipole-dipole and  exchange interactions as well as the  magnetic anisotropy. The geometry of the ferromagnetic sample is as follows: the length is 350 nm (along $ z $ axis), the width 50 nm (along $ y $ axis), and the thickness is  5 nm (along $ x $ axis).  In this case, the equilibrium magnetization  points in the in-plane along $ +z $ direction. The magnon temperature $ T_{F}^{m} = 5K $ and the temperature of the normal metal is set to zero $ T_N = 0 $. The time-dependent magnonic spin pumping currents $ I^x_s(t) $, $ I^y_s(t) $, and $ I^z_s(t) $ are plotted in Fig.\ref{figure2}. For  extended  samples, the dephasing of the transversal spin current components is faster and can hardly be captured through the micromagnetic simulations. As evident, the transversal components of the current oscillate randomly close to the zero value leading to the $\langle I_s^x(t)\rangle=0 $ and $ \langle I_s^y(t)\rangle=0 $. The longitudinal component  $ I_s^z(t) $ increases in time and saturates in the equilibrium regime.

Summarizing, we proposed  a theoretical approach to the time evolution of the spin Seebeck current. The approach is based on the time-dependent Fokker-Planck equation and supersymmetry arguments. We managed to derive the  analytical formula for the initial step of the buildup of the  spin current in the spin Seebeck effect. The  results are confirmed by full numerical calculations. The current experimental interest shows that ultrafast spin dynamics will play an increasingly significant role in spin caloritronics in the foreseeable future. Analytical tools for the time-dependent FP equation are quite limited.  Therefore, the alternative method proposed in our work should be useful for spin caloritronic studies.

\begin{acknowledgments}

This work is supported by the DFG through the SFB 762 and SFB-TRR 227, by Shota Rustaveli National Science Foundation of Georgia (SRNSFG) [Grant No. FR-19-4049], the National Research Center in Poland as a research project No. DEC-2017/27/B/ST3/02881, the National Natural Science
Foundation of China (Grants No. 11704415) and Natural Science Foundation of Hunan Province of China (Grant No. 2018JJ3629).
\end{acknowledgments}


\begin{thebibliography}{1}


\bibitem{Koopmans}
B. Koopmans, G. Malinowski, F. Dalla Longa, D. Steiauf, M.F\"ahnle, T. Roth, M. Cinchetti, and M. Aeschlimann, Nat. Mater. \textbf{9}, 259 (2010).
\bibitem{Tserkovnyak}
E. G. Tveten, A. Brataas, and Y. Tserkovnyak, Phys. Rev. B \textbf{92}, 180412 (2015).
\bibitem{Battiato}
M. Battiato, K. Carva, and P. M. Oppeneer, Phys.Rev. Lett.\textbf{105}, 027203 (2010).
\bibitem{Chubykalo-Fesenko}
U. Atxitia, O. Chubykalo-Fesenko, J. Walowski, A. Mann, and M. M\"unzenberg, Phys. Rev. B \textbf{81}, 174401 (2010).
\bibitem{Atxitia} U. Atxitia, Phys. Rev. B \textbf{98}, 014417 (2018);
E. Beaurepaire, J.-C. Merle, A. Daunois, and J.-Y. Bigot, Phys. Rev. Lett. \textbf{76}, 4250 (1996); A. Kirilyuk, A. V. Kimel, and T. Rasing, Rev. Mod. Phys. \textbf{82}, 2731 (2010).
\bibitem{xiguangfront}
X.-G. Wang, L. Chotorlishvili, and J. Berakdar
Front. Mater. 4, pp 19 (2017).
%
\bibitem{H. Risken} H. Risken,
The Fokker-Planck Equation: Methods of Solution and Applications (Springer Series in Synergetics) 2nd Edition, (1989).
\bibitem{W. T. Coffey}
W. T. Coffey and Y. P. Kalmykov, The Langevin Equation: with Applications to Stochastic Problems in Physics, Chemistry and Electrical Engineering vol 27 (Singapore: World Scientific),
(2012).
\bibitem{S. V. Titov}
D. J. Byrne, W. T. Coffey, W. J. Dowling, Y. P. Kalmykov, and S. V. Titov
Phys. Rev. B \textbf{93}, 064413 (2016).
\bibitem{D. A. Garanin1}
D. A. Garanin, Phys. Rev. B \textbf{55}, 3050 (1997).
\bibitem{D. A. Garanin2}
D. A. Garanin, Phys. Rev. B \textbf{98}, 144425 (2018).
\bibitem{miyazaki98}
K. Miyazaki and K. Seki, J. Chem. Phys. {\bf 108}, 7052 (1998).
\bibitem{Schwab}
L. Chotorlishvili, P. Schwab, Z. Toklikishvili, and J. Berakdar
Phys. Rev. B \textbf{82}, 014418 (2010).
\bibitem{Chotorlishvili2013}
L. Chotorlishvili, Z. Toklikishvili, V. K. Dugaev, J. Barna\'{s}, S. Trimper, and J. Berakdar
Phys. Rev. B \textbf{88}, 144429 (2013).
\bibitem{Chotorlishvili2019}
L. Chotorlishvili, Z. Toklikishvili, X.-G. Wang, V. K. Dugaev, J. Barna\'{s}, and J. Berakdar
Phys. Rev. B \textbf{99}, 024410 (2019).
\bibitem{XiaoBauerUchida}
J. Xiao, G. E. W. Bauer, K. C. Uchida, E. Saitoh, and S. Maekawa, Phys. Rev. B \textbf{81}, 214418 (2010).
\bibitem{K. I. Uchida} E. Saitoh and K. I. Uchida, Spin Seebeck effect, Spin
Current, Edited by Sadamichi Maekawa, Sergio O. Valen-
zuela, Eiji Saitoh, Takashi Kimura, Oxford University
press, (2012).
\bibitem{P. W. Brouwer}
T. S. Seifert, S. Jaiswal, J. Barker, S. T. Weber, I. Razdolski, J. Cramer, O. Gueckstock, S. F. Maehrlein, L. Nadvornik, S. Watanabe, C. Ciccarelli, A. Melnikov, G. Jakob, M. M\"unzenberg, S. T. B. Goennenwein, G. Woltersdorf, B. Rethfeld, P. W. Brouwer, M. Wolf, M. Kl\"aui, T. Kampfrath, Nature Communications \textbf{9}, 2899 (2018).

\bibitem{Tserkovnyak1} J. Foros, A. Brataas, Y. Tserkovnyak and G. E. Bauer, Phys. Rev. Lett. \textbf{95}, 016601 (2005).
\bibitem{Tserkovnyak2} Y. Tserkovniak, A. Brataas and G. E. W. Bauer, Phys. Rev. Lett. \textbf{88}, 117601 (2002).
\bibitem{AdachiUchidaSaitohMaekawa} H. Adachi, K. I. Uchida, E. Saitoh and S. Maekawa, Rep. Prog. Phys. \textbf{76}, 036501 (2013).
\bibitem{tsvelik}
M. V. Feigel'man and A. M. Tsvelik, Sov. Phys. JETP {\bf 56}, 823 (1982).
\bibitem{Melnikov} V. I. Mel\'nikov, Phys. Reports, \textbf{209}, 1(1991).
\bibitem{Witten}
E. Witten, Nucl. Phys. B \textbf{185}, 513(1981).
\bibitem{BernsteinBrown}
M. Bernstein and L. S. Brown, Phys. Rev. Lett. \textbf{52}, 1933 (1984).
\bibitem{Parisi}
G. Parisi, N. Sourlas, Phys. Rev. Lett. \textbf{43}, 744 (1979).
\bibitem{Sourlas}
G. Parisi, N. Sourlas, Nucl. Phys. B \textbf{206}, 321 (1982).
\bibitem{CaroliCaroli}
B. Caroli, C. Caroli B. Roulet and J. F. Gonyet, J. Stat. Phys. \textbf{22}, 515 (1980).
\bibitem{Girardello}
S. Cecotti, L. Girardello, Ann. Phys. \textbf{145}, 81 (1983).
\bibitem{Cecotti}
S. Cecotti, L. Girardello,  Nucl. Phys. B \textbf{239}, 573 (1984).
\bibitem{Horgan}
I. T. Drummond, R. R. Horgan, J. Phys. A \textbf{45}, 095005 (2012).
\bibitem{Kleinert}
H. Kleinert, S. V. Shabanov, Phys. Lett. A \textbf{235}, 105 (1997).
\bibitem{Khomenko}
A. I. Olemskoi, A. V. Khomenko, D. A. Olemskoi,  Phys. A \textbf{332}, 185 (2004).
\bibitem{Dijkgraaf}
R. Dijkgraaf, D. Orlando, S. Reffert, Nucl. Phys. B \textbf{824}, 365 (2010),
\bibitem{Ovchinnikov}
I. V. Ovchinnikov, Entropy \textbf{18}, 108 (2016).
\bibitem{Kunita}
H. Kunita, \textit{Stochastic Flows and Stochastic Differential Equations}, Cambridge University Press: Cambridge, UK (1997).
\bibitem{Baxendale}
P. H. Baxendale, S. V. Lototsky, \textit{Stochastic Differential Equations: Theory and Applications}, World Scientific, Singapore (2007).
\bibitem{Gardiner}
C. Gardiner, \textit{Stochastic Methods: A Handbook for theNatural and Social Sciences} Springer, Berlin (2009).
\bibitem{OstlerHinzkeNowak}
T. A. Ostler, M. O. A. Ellis, D. Hinzke, and U. Nowak, Phys. Rev. B \textbf{90}, 094402 (2014).
\bibitem{AbragamGoldman}
A. Abragam  M. Goldman \textit{Nuclear Magnetism: Order and Disorder}
Oxford University Press, (1982).
\end{thebibliography}
\end{document}